\documentclass[11pt,a4paper]{article}
\usepackage[T1]{fontenc}
\usepackage[utf8]{inputenc}
\usepackage{lmodern}
\usepackage[a4paper,margin=1in]{geometry}
\usepackage{microtype}
\usepackage{amsmath,amssymb}
\usepackage{graphicx}
\usepackage{booktabs,multirow,array}
\usepackage[super,sort,comma]{natbib}
\usepackage[section]{placeins}
\usepackage{caption}
\usepackage{float}
\usepackage[hidelinks]{hyperref}
\usepackage[all]{hypcap}

\makeatletter
\renewcommand\@biblabel[1]{$^{#1}$}
\makeatother
\title{DeepPV-MRF: End-to-End Partial-Volume Fraction Estimation from MR Fingerprinting Time Series}
\author{Tianyi Ding, Yang Gao, Zhuang Xiong, Feng Liu, Martijn A. Cloos, Jin Jin, Hongfu Sun}
\date{}
\hypersetup{
  pdftitle={DeepPV-MRF: End-to-End Partial-Volume Fraction Estimation from MR Fingerprinting Time Series},
  pdfauthor={Tianyi Ding; Yang Gao; Zhuang Xiong; Feng Liu; Martijn A. Cloos; Hongfu Sun},
  pdfkeywords={Magnetic Resonance Fingerprinting, Partial Volume Estimation, Deep Learning, Tissue Quantification}
}

\makeatletter
\renewcommand{\maketitle}{%
  \begin{center}\setlength{\parskip}{0pt}
    {\fontsize{18}{21}\selectfont\bfseries \@title\par}
    \vspace{10pt}
    {\normalsize
    Tianyi Ding$^{1}$, Yang Gao$^{1,2}$, Zhuang Xiong$^{3}$,\\[2pt]
    Feng Liu$^{1}$, Martijn A. Cloos$^{4,5}$,Jin Jin$^{6}$, Hongfu Sun$^{1,7}$\par}
    \vspace{8pt}
    {\small
    $^{1}$ School of Electrical Engineering and Computer Science, University of Queensland, QLD 4072, Australia\par
    $^{2}$ School of Computer Science and Engineering, Central South University, Changsha 410083, China\par
    $^{3}$ Image X Institute, Sydney School of Health Sciences, Faculty of Medicine and Health, University of Sydney, NSW 2006, Australia\par
    $^{4}$ Donders Institute for Brain Cognition and Behaviour, Radboud University, Nijmegen, Netherlands\par
    $^{5}$ Australian Institute for Bioengineering and Nanotechnology, University of Queensland, St Lucia, QLD, Australia\par
    $^{6}$ Siemens Healthineers Pty Ltd, Brisbane, Queensland, Australia\par
    $^{7}$ School of Engineering, University of Newcastle, Callaghan, NSW 2308, Australia\par}
    \vspace{6pt}
    {\small\textbf{Corresponding author:} Hongfu Sun\\
    School of Engineering, University of Newcastle, Callaghan, NSW 2308, Australia\\
    Email: \href{mailto:hongfu.sun@newcastle.edu.au}{\texttt{hongfu.sun@newcastle.edu.au}}\par}
  \end{center}
  \vspace{2pt}
}
\makeatother

\newcommand{\tableAllShots}{
  \begin{table*}[!t]
  \centering
  \caption{Performance comparison of different tissue fraction estimation methods under different shot numbers using simulated magnetic resonance fingerprinting (MRF) signals. Quantitative results are reported as mean \(\pm\) standard deviation across all synthetic test slices pooled from the 3 unseen test subjects.}
  \label{tab:pv_comparison_combined}

  \footnotesize
  \resizebox{\textwidth}{!}{
  \begin{tabular}{
  >{\centering\arraybackslash}p{0.4cm}
  >{\centering\arraybackslash}p{1.2cm}
  >{\centering\arraybackslash}p{2.3cm}
  >{\centering\arraybackslash}p{2.8cm}
  >{\centering\arraybackslash}p{2.4cm}
  >{\centering\arraybackslash}p{2.8cm}
  >{\centering\arraybackslash}p{2.4cm}
  }
  \toprule
   & \textbf{Tissue} & \textbf{Method} & \textbf{MAE ($\times10^3$)\,($\downarrow$)} & \textbf{PSNR\, ($\uparrow$)} & \textbf{SSIM\, ($\uparrow$)} & $\mathbf{r}_{\text{weighted}}$\, ($\uparrow$) \\
  \midrule

  \multirow{9}{*}{\rotatebox[origin=c]{90}{Single-Shot}}
  & \multirow{3}{*}{CSF}
    & DeepPV-MRF & \textbf{41.76 $\pm$ 5.39} & \textbf{30.81 $\pm$ 1.37} & \textbf{0.960 $\pm$ 0.014} & \textbf{0.957 $\pm$ 0.013} \\
  & & PV-DM & 128.07 $\pm$ 28.60 & 19.93 $\pm$ 1.72 & 0.643 $\pm$ 0.081 & 0.71 $\pm$ 0.103 \\
  & & PV-Pinv & 202.10 $\pm$ 35.96 & 15.22 $\pm$ 1.27 & 0.396 $\pm$ 0.076 & 0.338 $\pm$ 0.314 \\
\cmidrule(lr){3-7}

  & \multirow{3}{*}{WM}
    & DeepPV-MRF & \textbf{74.39 $\pm$ 10.65} & \textbf{24.80 $\pm$ 1.38} & \textbf{0.916 $\pm$ 0.018} & \textbf{0.846 $\pm$ 0.034} \\
  & & PV-DM & 244.25 $\pm$ 35.03 & 12.57 $\pm$ 1.22 & 0.527 $\pm$ 0.039 & 0.414 $\pm$ 0.096 \\
  & & PV-Pinv & 279.27 $\pm$ 26.89 & 11.76 $\pm$ 0.79 & 0.448 $\pm$ 0.047 & 0.346 $\pm$ 0.105 \\
  \cmidrule(lr){3-7}
  & \multirow{3}{*}{GM}
    & DeepPV-MRF & \textbf{75.69 $\pm$ 9.41} & \textbf{23.84 $\pm$ 1.33} & \textbf{0.922 $\pm$ 0.014} & \textbf{0.879 $\pm$ 0.026} \\
  & & PV-DM & 295.82 $\pm$ 54.10 & 11.51 $\pm$ 1.46 & 0.525 $\pm$ 0.073 & 0.337 $\pm$ 0.116 \\
  & & PV-Pinv & 263.02 $\pm$ 28.97 & 12.03 $\pm$ 0.67 & 0.47 $\pm$ 0.066 & 0.325 $\pm$ 0.138 \\
  \midrule
  \multirow{9}{*}{\rotatebox[origin=c]{90}{Three-Shot}}
  & \multirow{3}{*}{CSF}
    & DeepPV-MRF & \textbf{25.26 $\pm$ 6.26} & \textbf{34.48 $\pm$ 2.04} & \textbf{0.978 $\pm$ 0.011} & \textbf{0.981 $\pm$ 0.007} \\
  & & PV-DM & 113.91 $\pm$ 28.91 & 21.17 $\pm$ 1.89 & 0.694 $\pm$ 0.092 & 0.772 $\pm$ 0.096 \\
  & & PV-Pinv & 173.54 $\pm$ 18.62 & 16.27 $\pm$ 0.87 & 0.462 $\pm$ 0.047 & 0.574 $\pm$ 0.133 \\
  \cmidrule(lr){3-7}
  & \multirow{3}{*}{WM}
    & DeepPV-MRF & \textbf{53.42 $\pm$ 8.74} & \textbf{27.69 $\pm$ 1.38} & \textbf{0.947 $\pm$ 0.011} & \textbf{0.924 $\pm$ 0.016} \\
  & & PV-DM & 204.97 $\pm$ 37.88 & 14.02 $\pm$ 1.59 & 0.586 $\pm$ 0.045 & 0.557 $\pm$ 0.127 \\
  & & PV-Pinv & 266.49 $\pm$ 37.60 & 12.28 $\pm$ 1.25 & 0.472 $\pm$ 0.058 & 0.409 $\pm$ 0.137 \\
  \cmidrule(lr){3-7}
  & \multirow{3}{*}{GM}
    & DeepPV-MRF & \textbf{52.63 $\pm$ 9.61} & \textbf{26.64 $\pm$ 1.46} & \textbf{0.946 $\pm$ 0.013} & \textbf{0.935 $\pm$ 0.019} \\
  & & PV-DM & 256.91 $\pm$ 63.73 & 12.86 $\pm$ 1.99 & 0.592 $\pm$ 0.089 & 0.431 $\pm$ 0.149 \\
  & & PV-Pinv & 272.08 $\pm$ 40.94 & 12.10 $\pm$ 1.13 & 0.451 $\pm$ 0.090 & 0.297 $\pm$ 0.180 \\
  \midrule
  \multirow{9}{*}{\rotatebox[origin=c]{90}{Six-Shot}}
  & \multirow{3}{*}{CSF}
    & DeepPV-MRF & \textbf{22.93 $\pm$ 4.93} & \textbf{35.12 $\pm$ 1.80} & \textbf{0.980 $\pm$ 0.010} & \textbf{0.984 $\pm$ 0.007} \\
  & & PV-DM & 110.91 $\pm$ 29.51 & 21.41 $\pm$ 1.95 & 0.709 $\pm$ 0.094 & 0.782 $\pm$ 0.096 \\
  & & PV-Pinv & 171.64 $\pm$ 19.55 & 16.34 $\pm$ 0.92 & 0.468 $\pm$ 0.050 & 0.584 $\pm$ 0.136 \\
  \cmidrule(lr){3-7}
  & \multirow{3}{*}{WM}
    & DeepPV-MRF & \textbf{43.16 $\pm$ 6.25} & \textbf{29.35 $\pm$ 1.39} & \textbf{0.954 $\pm$ 0.011} & \textbf{0.946 $\pm$ 0.012} \\
  & & PV-DM & 194.39 $\pm$ 36.41 & 14.44 $\pm$ 1.59 & 0.606 $\pm$ 0.042 & 0.601 $\pm$ 0.120 \\
  & & PV-Pinv & 263.58 $\pm$ 39.68 & 12.41 $\pm$ 1.36 & 0.482 $\pm$ 0.061 & 0.430 $\pm$ 0.145 \\
  \cmidrule(lr){3-7}
  & \multirow{3}{*}{GM}
    & DeepPV-MRF & \textbf{43.96 $\pm$ 6.97} & \textbf{28.22 $\pm$ 1.36} & \textbf{0.958 $\pm$ 0.011} & \textbf{0.954 $\pm$ 0.013} \\
  & & PV-DM & 246.49 $\pm$ 63.81 & 13.25 $\pm$ 2.02 & 0.613 $\pm$ 0.088 & 0.456 $\pm$ 0.150 \\
  & & PV-Pinv & 270.98 $\pm$ 44.04 & 12.18 $\pm$ 1.23 & 0.455 $\pm$ 0.096 & 0.308 $\pm$ 0.192 \\
  \bottomrule
  \end{tabular}
  }
  
  \vspace{2mm}
  \footnotesize
  Arrows ($\uparrow$/$\downarrow$) indicate whether higher or lower values are better for each metric. Bold indicates the best performance per tissue and metric.
  \end{table*}
}

\newcommand{\tableInvivoCrossShotConsistency}{
\begin{table}[!t]
\centering
\caption{Cross-shot consistency of DeepPV-MRF in the three \textit{in vivo} volunteers, using the six-shot prediction as an internal high-shot reference. Lower mean absolute error (MAE) indicates greater agreement between the lower-shot and six-shot predictions. The six-shot prediction is used only as an internal consistency reference and should not be interpreted as \textit{in vivo} ground truth.}
\label{tab:invivo_cross_shot_consistency}
\footnotesize
\begin{tabular}{llccc}
\toprule
Tissue & Subject & 1-shot vs. 6-shot & 3-shot vs. 6-shot & Improvement (\%) \\
\midrule
\multirow{3}{*}{CSF} & sub01 & 0.164 & 0.138 & 15.9 \\
 & sub02 & 0.131 & 0.126 & 3.5 \\
 & sub03 & 0.130 & 0.109 & 16.5 \\
\cmidrule(lr){1-5}
\multirow{3}{*}{WM} & sub01 & 0.245 & 0.179 & 27.1 \\
 & sub02 & 0.227 & 0.194 & 14.8 \\
 & sub03 & 0.191 & 0.181 & 5.1 \\
\cmidrule(lr){1-5}
\multirow{3}{*}{GM} & sub01 & 0.317 & 0.226 & 28.7 \\
 & sub02 & 0.286 & 0.247 & 13.6 \\
 & sub03 & 0.256 & 0.225 & 12.1 \\
\bottomrule
\end{tabular}
\par\vspace{1mm}
\begin{minipage}{0.95\columnwidth}
\footnotesize
Improvement denotes the relative MAE reduction from 1-shot vs. 6-shot to 3-shot vs. 6-shot.
\end{minipage}
\end{table}
}

\newcommand{\figDataGen}{
  \begin{figure}[!t]
    \centering
    \includegraphics[width=0.90\columnwidth]{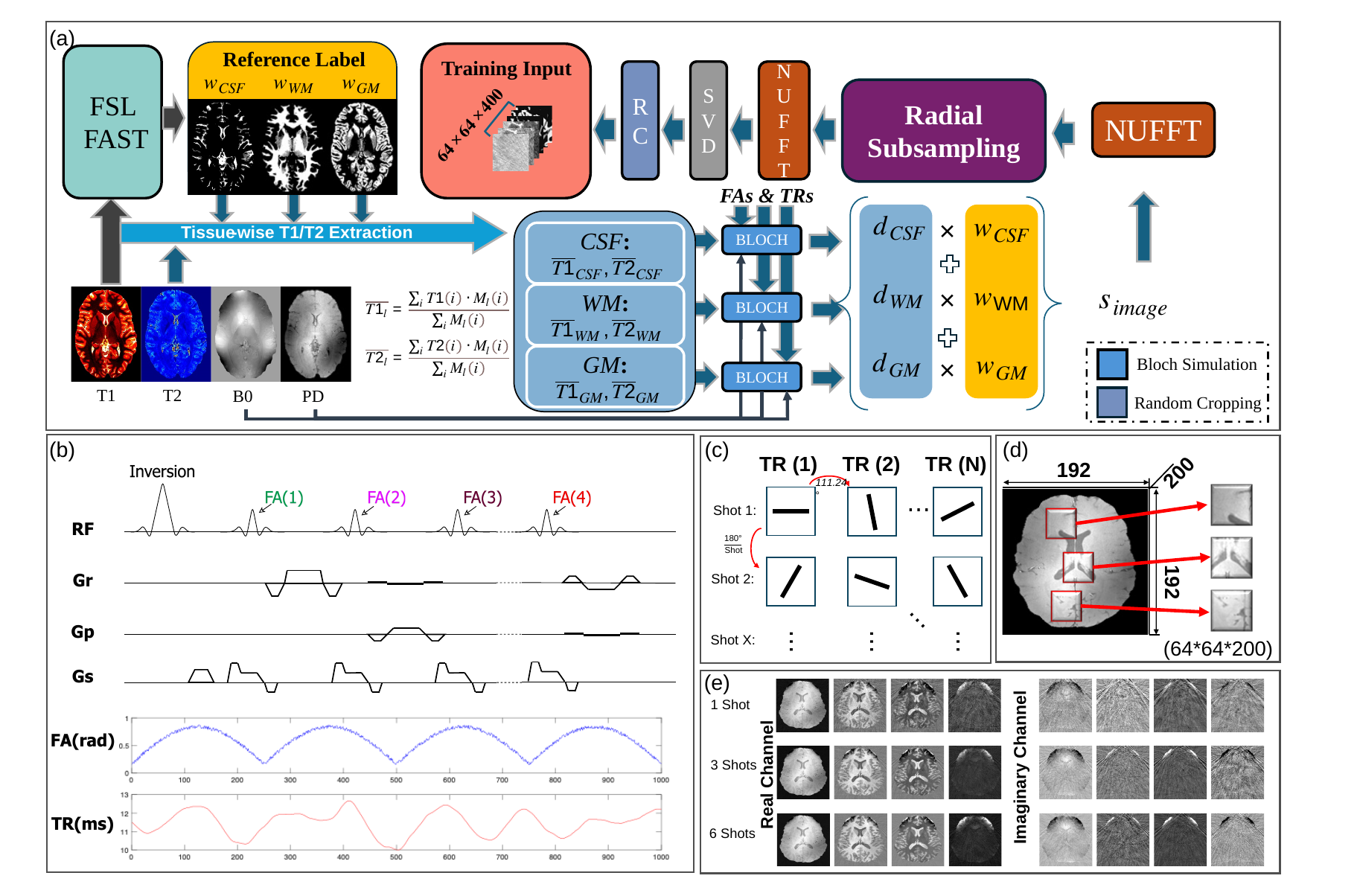}
    \caption{(a) Synthetic data generation pipeline, (b) inversion recovery balanced steady-state free precession sequence, (c) radial subsampling trajectory, (d) random cropping, and (e) examples of the first four real and imaginary channels of the training input data.}
    \label{fig:datagen}
  \end{figure}
}

\newcommand{\figInvivoNew}{
  \begin{figure*}[htbp]
    \centering
    \includegraphics[width=0.9\textwidth]{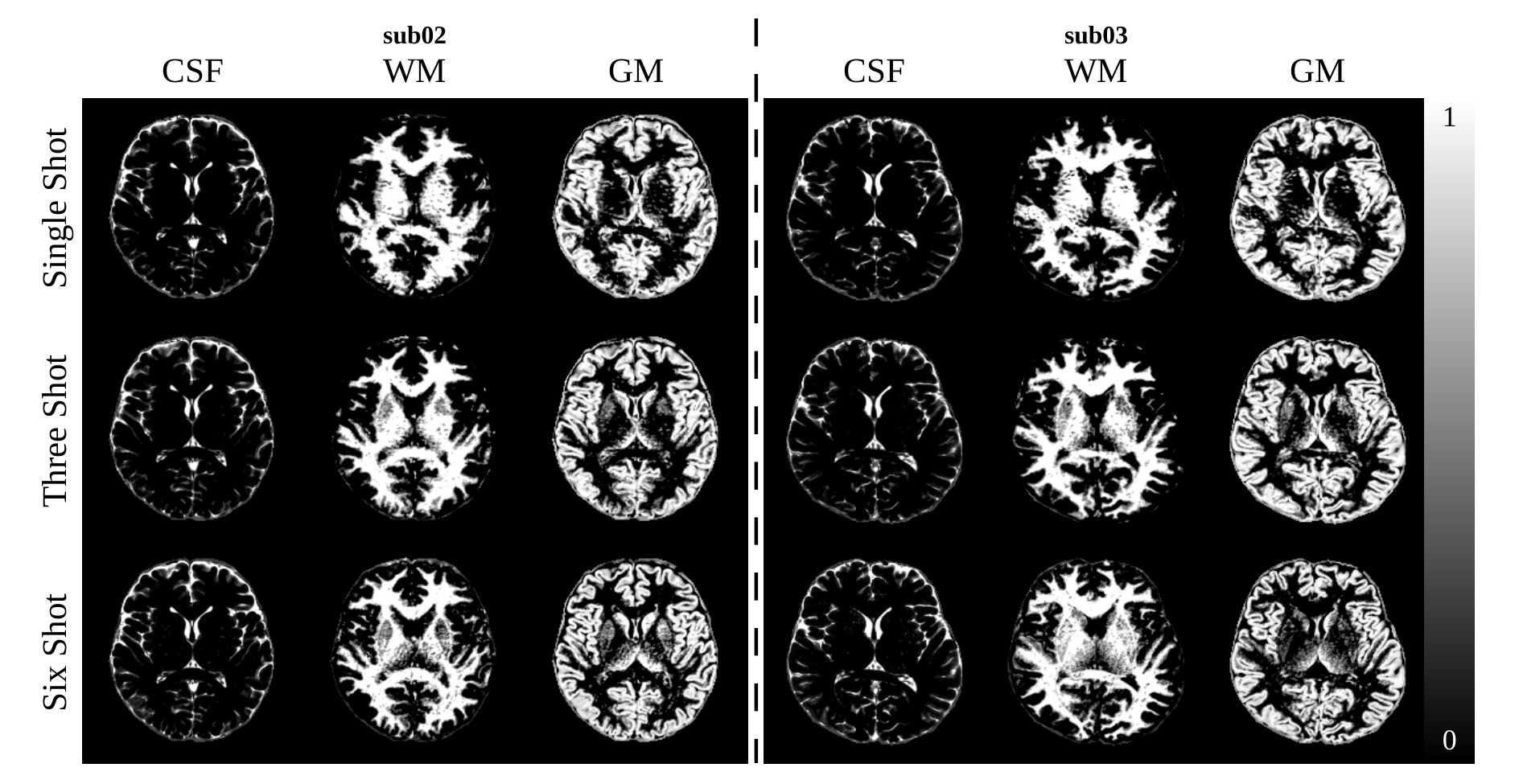}
    \caption{Representative \textit{in vivo} partial-volume maps from two volunteers (sub 02 and sub 03) reconstructed using the proposed DeepPV-MRF model under 1-shot, 3-shot, and 6-shot acquisitions. For each volunteer, the same anatomical slice is shown across shot settings to qualitatively assess consistency across acquisition schemes. Separate maps are shown for CSF, WM, and GM tissue fractions. All maps are displayed on the same scale from 0 to 1.}
    \label{fig:invivonew}
  \end{figure*}
}

\newcommand{\figInvivoSNR}{
  \begin{figure*}[htbp]
    \centering
    \includegraphics[width=1\textwidth]{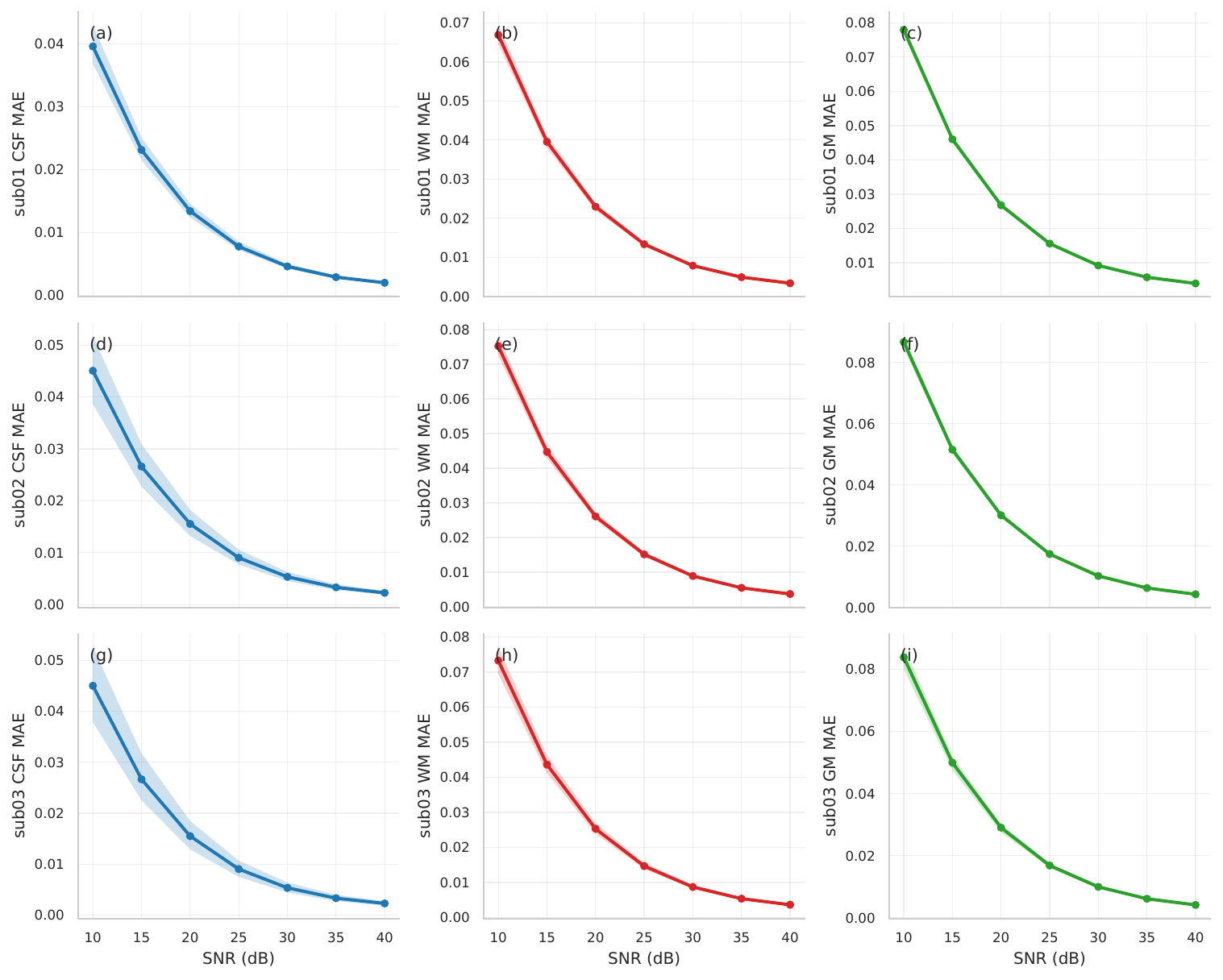}
    \caption{\textit{In vivo} noise-sensitivity analysis for the single-shot setting across the three volunteers. The input MRF signals were perturbed with complex Gaussian noise over signal-to-noise ratio (SNR) levels from 10~dB to 40~dB. Rows correspond to sub01, sub02, and sub03, and columns show cerebrospinal fluid (CSF), white matter (WM), and gray matter (GM). The plotted values represent the mean absolute error (MAE) between predictions from noisy inputs and the corresponding clean-input predictions. Solid lines indicate the mean across slices, and shaded regions denote the 95\% confidence intervals.}
    \label{fig:invivosnr}
  \end{figure*}
}

\newcommand{\figNetWork}{
  \begin{figure}[!t]  
    \centering
    \includegraphics[width=0.88\columnwidth]{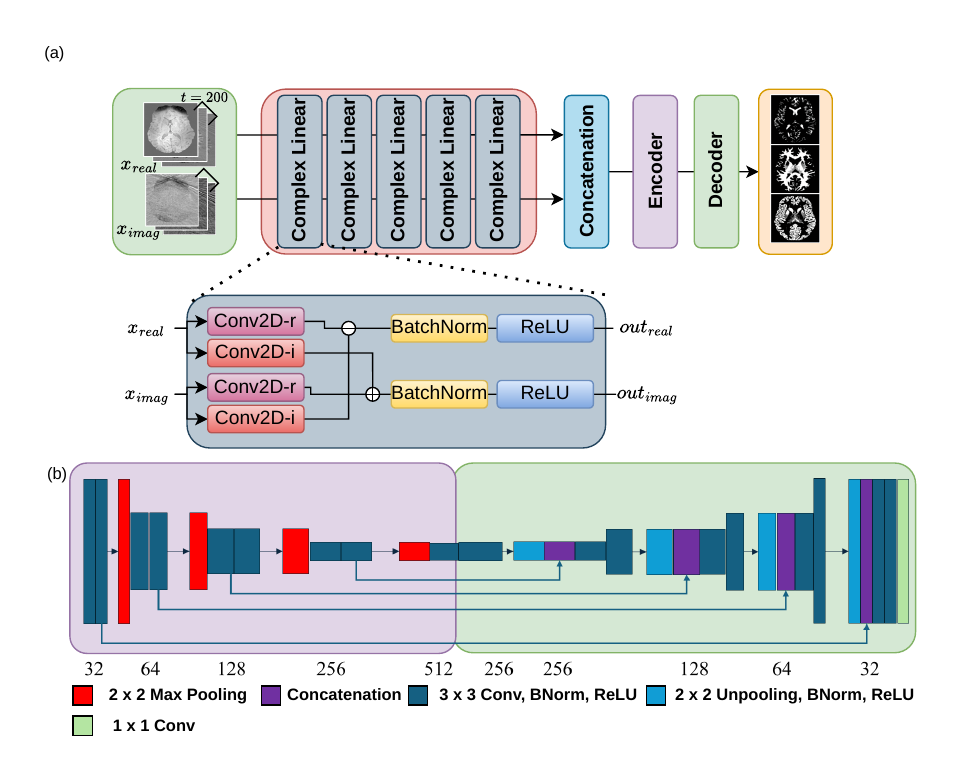} 
    \caption{(a) Overview of the DeepPV-MRF framework, including real and imaginary input channels and tissue-fraction (TF) prediction. 
    (b) U-Net encoder structure with channel depths and block functions (left), and decoder structure for TF prediction (right).}
\label{fig:network}

  \end{figure}
}

\newcommand{\figSimulation}{
  \begin{figure*}[!t]
    \centering
    \includegraphics[width=1.0\textwidth]{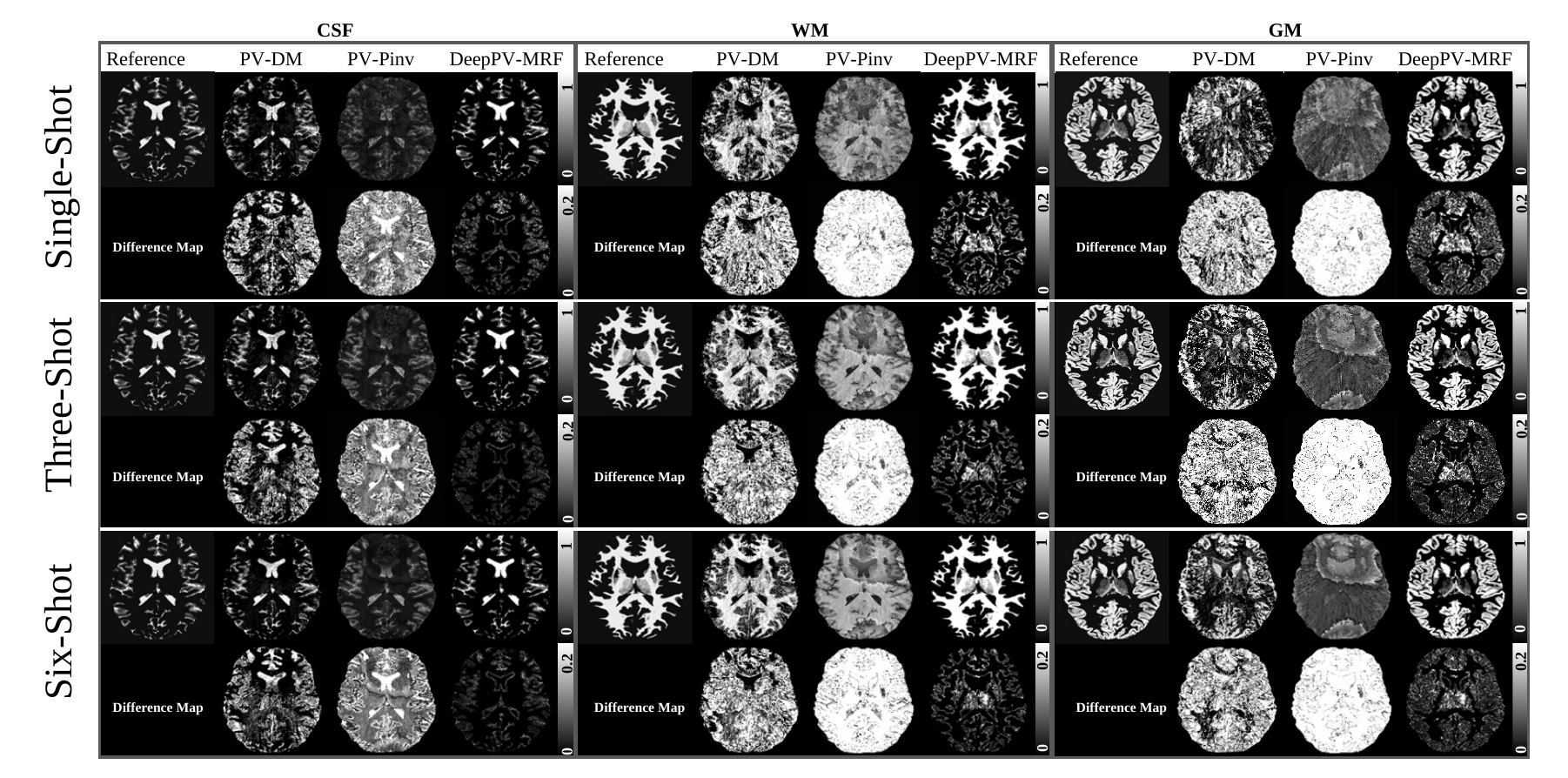}
       \caption{
    Qualitative comparison of tissue fraction maps for cerebrospinal fluid (CSF), white matter (WM), and gray matter (GM) under single-shot, three-shot, and six-shot acquisitions in the synthetic test set. For each tissue type and acquisition setting, four methods are compared: Reference Label, partial volume dictionary matching (PV-DM), partial volume pseudoinverse (PV-Pinv), and the proposed DeepPV-MRF. Corresponding absolute difference maps are shown in the second row for each acquisition setting to visualize voxel-wise reconstruction errors. All maps are displayed using a common intensity window of [0, 1] for tissue fractions and [0, 0.2] for difference maps.
    }

    \label{fig:simulation}
  \end{figure*}
}

\newcommand{\figScatterPlot}{
  \begin{figure*}[!t]
    \centering
    \includegraphics[width=0.8\textwidth]{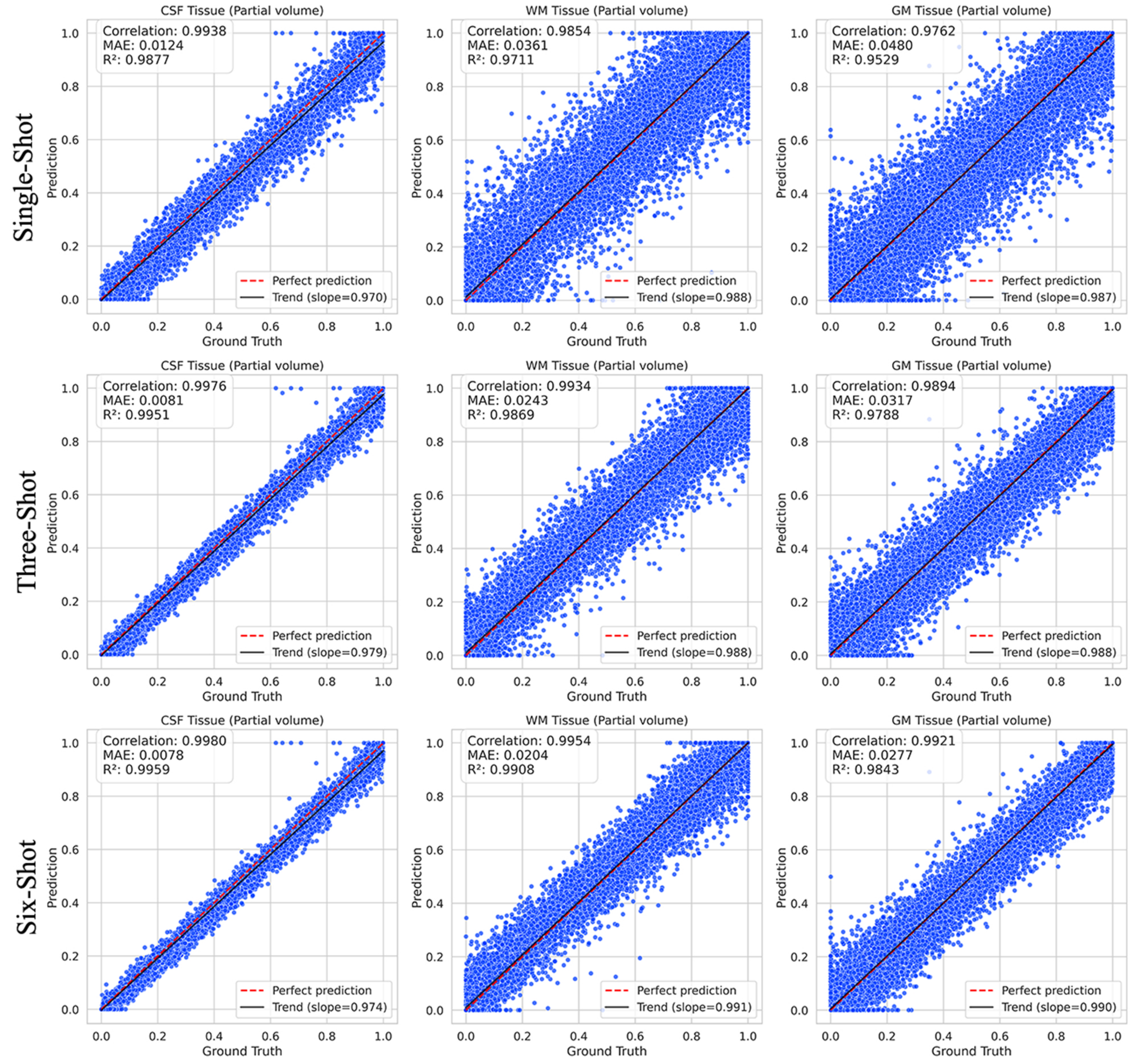}
    \caption{
    Scatter plots comparing DeepPV-MRF predicted tissue fraction values against reference FSL-FAST labels for cerebrospinal fluid (CSF), white matter (WM), and gray matter (GM) across single-shot (top row), three-shot (middle row), and six-shot (bottom row) acquisition settings. Each plot displays voxel-wise predictions (y-axis) versus corresponding reference values (x-axis). Quantitative metrics include Pearson correlation coefficient, MAE, coefficient of determination (\(R^2\)), and the slope of the fitted regression line. The red dashed line indicates the line of perfect agreement.
    }
    
    \label{fig:scatterplot}
  \end{figure*}
}
\newcommand{\figInvivoOne}{
  \begin{figure*}[!t]
    \centering
    \includegraphics[width=1.0\textwidth]{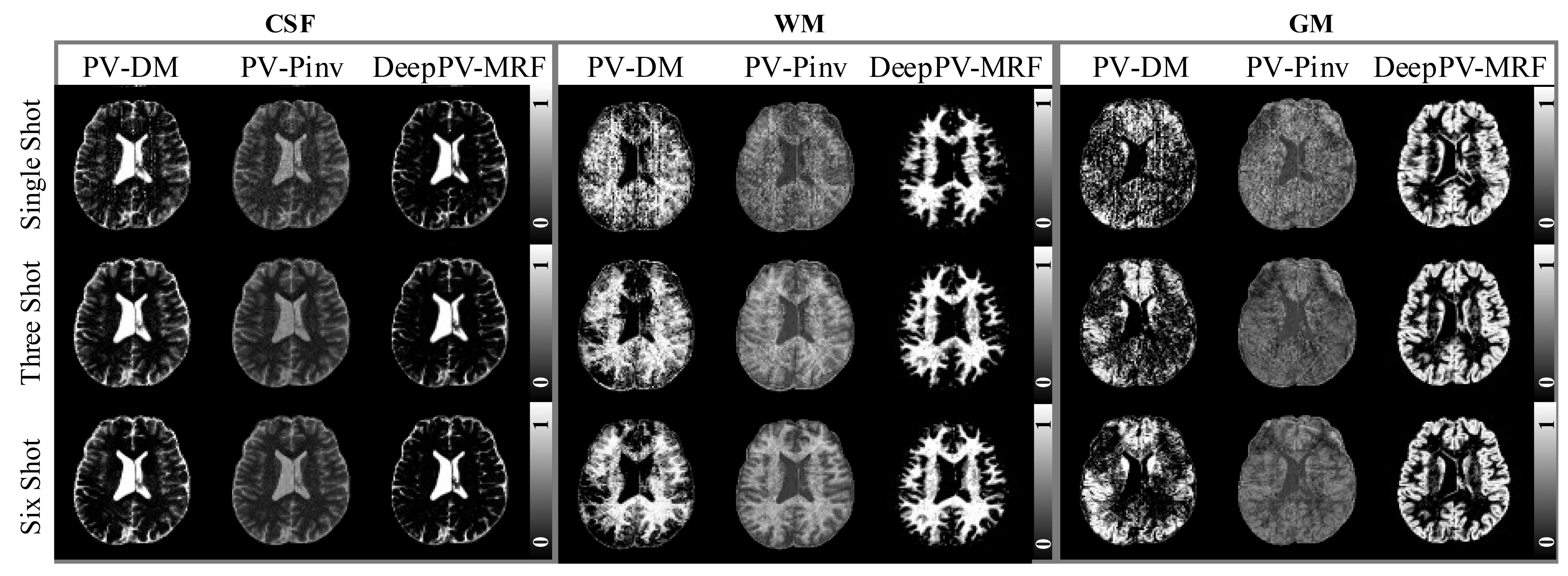}
   \caption{
    Comparison of predicted tissue fraction maps for cerebrospinal fluid (CSF), white matter (WM), and gray matter (GM) across single-shot, three-shot, and six-shot \textit{in vivo} MRF acquisitions for sub 01. For each tissue type, results from three methods—partial volume dictionary matching (PV-DM), partial volume pseudoinverse (PV-Pinv), and the proposed DeepPV-MRF—are shown from left to right. Each row corresponds to a different acquisition setting, with intensity values mapped from 0 to 1. 
    }

    \label{fig:invivo1}
  \end{figure*}
}

\newcommand{\figMixer}{
  \begin{figure*}[t]
    \centering
    \includegraphics[width=0.9\textwidth]{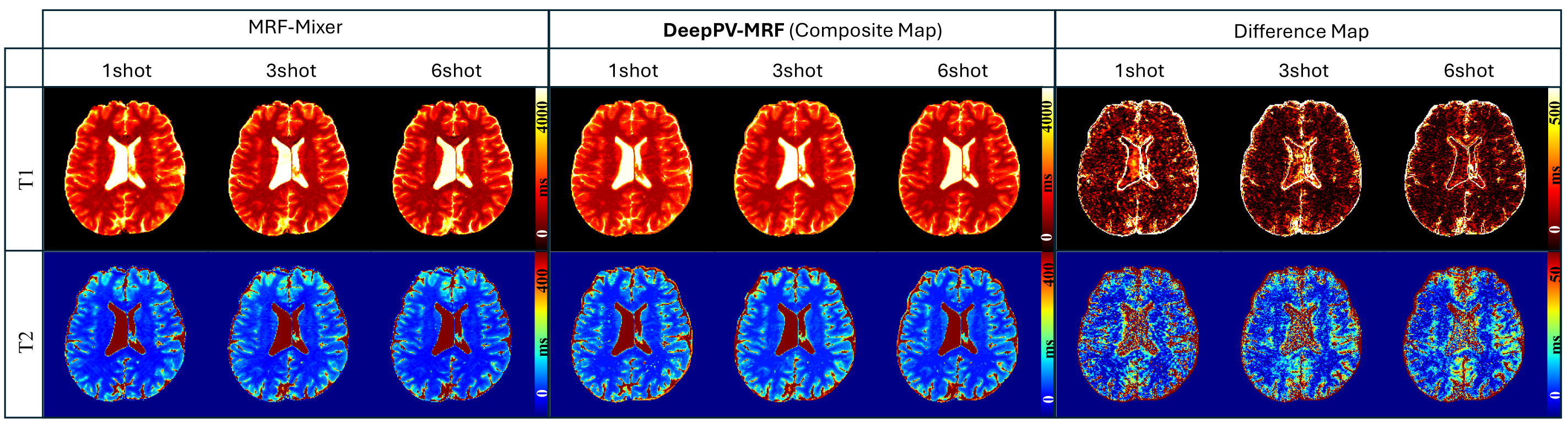}
    \caption{ \textit{in vivo} comparison of T1 and T2 maps from MRF-Mixer (left) and composite maps derived from DeepPV-MRF (middle), with their absolute difference shown on the right. The composite maps combine estimated tissue fractions with pure-tissue relaxation values. The difference maps confirm that the largest discrepancies are concentrated at tissue boundaries, consistent with partial volume effects. }

    \label{fig:testmixer}
  \end{figure*}
}

\newcommand{\tableAblation}{
\begin{table*}[!t]
\centering
\caption{Ablation study on the external 1-shot simulation test set. Mean absolute error (MAE) denotes tissue-wise $\mathrm{MAE}_{\mathrm{tissue}}$ and is reported in $\times 10^3$. Macro-averaged Peak Signal-to-Noise Ratio (PSNR) and Structural Similarity Index Measure (SSIM) are unweighted averages across cerebrospinal fluid (CSF), white matter (WM), and gray matter (GM). The best value for each metric within each ablation family is shown in bold, with lower MAE and higher PSNR/SSIM indicating better performance. The full model denotes the architecture used for the main results; the B0/PD-head row adds auxiliary heads to this model.}
\label{tab:ablation}
\footnotesize
\resizebox{\textwidth}{!}{%
\begin{tabular}{llcccccc}
\toprule
Family & Setting & CSF MAE & WM MAE & GM MAE & Avg. MAE & Avg. PSNR & Avg. SSIM \\
\midrule
\multicolumn{8}{l}{\textit{Core ablations}} \\
& \textbf{Full model} & \textbf{41.76} & \textbf{74.39} & \textbf{75.69} & \textbf{63.95} & \textbf{26.48} & \textbf{0.933} \\
& With B0/PD heads & 43.42 & 77.70 & 78.51 & 66.55 & 26.14 & 0.922 \\
& Temporal encoder only & 52.12 & 110.25 & 106.26 & 89.54 & 23.95 & 0.890 \\
& No spatial gradient loss & 44.45 & 77.69 & 79.55 & 67.23 & 26.05 & 0.925 \\
\midrule
\multicolumn{8}{l}{\textit{Latent dimension sweep}} \\
& Latent dim = 8 & 41.10 & \textbf{73.62} & \textbf{75.11} & \textbf{63.28} & \textbf{26.61} & \textbf{0.933} \\
& Latent dim = 16 & 41.76 & 74.39 & 75.69 & 63.95 & 26.48 & 0.933 \\
& Latent dim = 32 & \textbf{38.32} & 84.34 & 80.72 & 67.79 & 25.92 & 0.919 \\
& Latent dim = 64 & 42.13 & 88.46 & 85.54 & 72.05 & 25.53 & 0.911 \\
& Latent dim = 128 & 43.41 & 87.44 & 85.54 & 72.13 & 25.63 & 0.911 \\
\midrule
\multicolumn{8}{l}{\textit{Spatial regularization sweep}} \\
& $\lambda_{\text{spatial}} = 0.01$ & 40.89 & 76.35 & 76.94 & 64.73 & 26.45 & 0.932 \\
& $\lambda_{\text{spatial}} = 0.05$ & 41.76 & \textbf{74.39} & \textbf{75.69} & \textbf{63.95} & \textbf{26.48} & \textbf{0.933} \\
& $\lambda_{\text{spatial}} = 0.10$ & \textbf{39.22} & 78.75 & 77.23 & 65.07 & 26.20 & 0.925 \\
& $\lambda_{\text{spatial}} = 0.50$ & 41.29 & 76.25 & 77.13 & 64.89 & 26.38 & 0.929 \\
\bottomrule
\end{tabular}%
}
\end{table*}
}

\begin{document}
\maketitle

\begin{abstract}
{\bf Background:} Accurate partial volume (PV) estimation, i.e., subvoxel tissue fractions, from Magnetic Resonance Fingerprinting (MRF) remains challenging, as conventional methods such as dictionary matching or segmentation tools like FSL FAST rely on multi-stage iterative pipelines, where inaccuracies in early steps (e.g., parametric map estimation) can propagate and compromise the final tissue fraction maps.

{\bf Purpose:}  This study introduces DeepPV-MRF, a deep learning framework designed to directly map raw MRF signals to voxel-wise tissue fraction maps. The goal is to provide a more accurate and efficient alternative to existing iterative and dictionary-based PV estimation methods.

{\bf Methods:} DeepPV-MRF employs a cascaded spatial-temporal neural network trained on subject-derived simulations generated through Bloch modeling with radial k-space undersampling. The   integrates voxel-wise temporal encoding with spatial U-Net refinement to estimate CSF, GM, and WM fractions without intermediate T1/T2 maps. Performance was evaluated against PV dictionary matching (PV-DM), and PV pseudoinverse (PV-Pinv), using both simulated and \textit{in vivo} data.

{\bf Results:} The proposed method substantially outperformed competing methods across all acquisition settings. In single-shot simulations, the CSF/WM/GM MAEs were 0.0418/0.0744/0.0757, compared with 0.1281/0.2443/0.2958 for PV-DM and 0.2021/0.2793/0.2630 for PV-Pinv; the corresponding SSIM values were 0.960/0.916/0.922. The \textit{in vivo} experiments qualitatively showed sharper anatomical boundaries and smoother tissue transitions, while the cross-shot analysis supported improved consistency across volunteers.

{\bf Conclusions:} In conclusion, DeepPV-MRF offers a robust, accurate, and efficient framework for direct partial volume estimation in MRF. By eliminating reliance on intermediate reconstructions and improving performance particularly in low-shot scenarios, the method addresses critical weaknesses of existing approaches. These results support the practical feasibility of DeepPV-MRF and motivate further validation before clinical deployment.

\end{abstract}

\clearpage

\section{Introduction}

\label{sec:introduction}
Quantitative MRI (qMRI) is a non-invasive imaging framework that enables numerical characterization of tissue microstructure \cite{tohka2014partial}. Broadly, qMRI applications fall into two major categories: parametric mapping and tissue composition estimation. The former focuses on extracting intrinsic tissue properties such as longitudinal relaxation time (\(T_1\)) and transverse relaxation time (\(T_2\)), which serve as biophysical biomarkers for disease diagnosis. The latter aims to estimate metrics such as volume or thickness of tissue such as gray matter (GM), white matter (WM), and cerebrospinal fluid (CSF) in the brain. 

While parametric maps provide valuable insights into tissue biophysics, they do not directly account for tissue mixing within a voxel \cite{plante1991sources}. Conventionally, tissue composition estimation assumes that each voxel contains a single homogeneous tissue type \cite{deshmane2019partial}. However, this assumption fails in regions where voxels span across anatomical boundaries. This limitation becomes particularly important in clinical brain imaging, where partial volume (PV) effects are prevalent due to the limited spatial resolution of MRF \cite{bedell1998volumetric,shenton2001review,weiner2013alzheimer}. PV effects can substantially impair the accuracy of tissue quantification \cite{plante1991sources}, lesion size estimations \cite{kikinis1999quantitative}, and brain atrophy assessment \cite{gao2014sub}.

To characterize sub-voxel tissue heterogeneity, voxel-wise tissue fraction estimation—i.e., quantifying the relative proportions of GM, WM, and CSF in each voxel—is essential. Traditional approaches to PV estimation typically rely on regularized statistical models, such as Gaussian mixture models (GMMs) combined with expectation-maximization (EM) \cite{wells1996adaptive,tohka2004fast,cuadra2005comparison,chappell2011partial}, often applied to multi-contrast structural MRI. These methods assume that each tissue class exhibits a well-defined intensity distribution across modalities and that voxel intensities result from a linear mixture of pure tissue responses. While effective in structural imaging, these assumptions do not hold in quantitative techniques like Magnetic Resonance Fingerprinting (MRF), where voxel signals represent non-linear temporal evolutions rather than static intensities.

MRF \cite{ma2013magnetic} signal representation captures multi-parametric information in a single scan, making it an attractive candidate for direct tissue composition estimation. One early attempt to incorporate PV modeling into MRF is the partial volume dictionary matching (PV-DM) method \cite{ma2013magnetic,deshmane2019partial}, which estimates tissue fractions through dictionary matching. While promising, this method inherits several limitations from the underlying dictionary-based paradigm. First, dictionary-based MRF reconstruction becomes computationally demanding as the size and resolution of the dictionary increase, especially for applications involving partial volume modeling where dense sampling of tissue parameter space may be necessary \cite{ma2013magnetic,mcgivney2014svd,yang2018low,poorman2020magnetic}. Second, PV-DM relies on a strong “pure-tissue” assumption, requiring manually defined reference clusters with fixed relaxation times for GM, WM, and CSF. This introduces subjectivity and operational complexity, making it difficult to deploy in automated or clinical pipelines. These challenges motivate the need for more flexible, data-driven approaches to sub-voxel tissue fraction estimation in the MRF context.

Recent studies have explored deep learning as an alternative to dictionary-based MRF reconstruction, demonstrating improved efficiency and scalability \cite{cohen2018mr,fang2019deep,golbabaee2019geometry,cao2020development,li2024deep,li2024enhancing,li2025deep,ding2025mrf}. These methods typically train convolutional or transformer-based models to map MRF time-series signals to voxel-wise tissue parameters such as \(T_1\) and \(T_2\). For example, DRONE \cite{cohen2018mr} first introduced a fully connected neural network to directly regress quantitative maps from signal evolutions. SCQ \cite{fang2019deep} extended this approach using convolutional architectures. More recent transformer-based models \cite{li2024deep} have shown further improvements in spatial consistency and generalization. However, most existing models are designed with a single-compartment model and do not explicitly estimate tissue fractions in partial volume voxels. While these models have demonstrated strong performance in parametric map reconstruction, they are not designed to resolve sub-voxel tissue mixing. As a result, partial volume effects remain unaddressed in most existing MRF-based deep learning pipelines, limiting their utility in tasks that require accurate tissue composition analysis. This limitation motivates the need for models capable of explicitly estimating sub-voxel tissue fractions from MRF signals.

To this end, we present DeepPV-MRF, an end-to-end deep learning framework for directly estimating voxel-wise tissue fraction maps from aliased MRF signal evolutions. The goal is to estimate the relative fractions of CSF, GM, and WM within each voxel, without relying on intermediate \(T_1\) or \(T_2\) maps. Extending the methodology proposed in MRF-Mixer~\cite{ding2025mrf}, we construct a fully simulated training dataset in which each voxel is assigned reference tissue fractions derived from high-resolution anatomical segmentations and corresponding pure-tissue relaxometry estimated from \textit{in vivo} acquisitions. Unlike conventional partial volume estimation techniques such as partial volume pseudoinverse (PV-Pinv) and PV-DM~\cite{deshmane2019partial}, which depend on predefined pure tissue fingerprints and manually designed subdictionaries, our method learns the mapping from MRF signals to tissue fractions directly from data, removing the need for dictionary construction, signal matching, or parameter discretization.

\section{Related Work}

\subsection{Magnetic Resonance Fingerprinting Parameter Mapping}

MRF represents an approach to MRI data acquisition and analysis that enables simultaneous quantification of multiple tissue parameters through a single abbreviated scan \cite{ma2013magnetic}. During the dictionary matching (DM) process, the  signal evolution observed in each voxel \(i\), termed the \textit{tissue fingerprint}, is represented as \(s_i \in \mathbb{C}^{1 \times t}\), comprising \(t\) time points. This fingerprint is systematically compared against a precomputed dictionary \(D_{\text{MRF}} \in \mathbb{C}^{t \times n}\), where each column \(d_j \in \mathbb{C}^{t}\) represents a signal evolution simulated from a unique combination of tissue parameters \(\{T_1, T_2\}\). The best-matching dictionary entry \(d_{i,\text{MRF match}}\) is identified as the one that maximizes the normalized dot product magnitude with the observed signal \cite{deshmane2019partial}. Mathematically, this matching procedure can be expressed as:
\begin{equation}
d_{i,\text{MRF match}} = \arg \max_{j=1,\dots,n} \left| \langle s_i, d_j \rangle \right|.
\end{equation}

The tissue parameters corresponding to the selected dictionary entry are then assigned to voxel \(i\). To accelerate the DM process, many approaches have been proposed  \cite{mcgivney2014svd,yang2018low,wang2020snapmrf,cauley2015fast}.Recently, deep learning methodologies have been proposed to bypass the DM process altogether \cite{cohen2018mr,fang2019deep,golbabaee2019geometry,li2024deep,li2025deep,ding2025mrf}. However, these deep learning approaches typically do not address the partial volume estimation.

\subsection{Partial Volume Estimation}

Conventional approaches to addressing the PV effect typically model the voxel intensity as a linear combination of signals from constituent tissue types, weighted by their fractional volumes within the voxel \cite{choi1991partial,tohka2014partial}. Mathematically, this relationship is described by:

\begin{equation}
X_{i} = \sum_{k=1}^{K} f_{i,k} x_{i,k}, \quad \text{subject to} \quad \sum_{k=1}^{K} f_{i,k} = 1,
\end{equation}

where \(X_{i}\) denotes the observed intensity at voxel \(i\), \(x_{i,k}\) represents the pure intensity of tissue type \(k\) at voxel \(i\), \(f_{i,k}\) denotes the fractional volume of tissue type \(k\) within voxel \(i\), and \(K\) indicates the total number of tissue types present.

Given a sufficient number of independent data channels, fractional tissue volumes can be directly estimated by solving this linear system via a least squares approach.

\subsection{Integrating Partial Volume Estimation with MRF}
Within the MRF framework, PV estimation has been approached by modeling each voxel's signal as a linear combination of pure tissue fingerprints. Let \(s_i\) denote the MRF signal at voxel \(i\), and let \(D_{sub} \in \mathbb{C}^{t \times m}\) represent a subdictionary of signal evolutions for \(m\) tissue types. Then, the signal can be modeled as:

\begin{equation}
s_i = D_{sub} \mathbf{w}_i,
\end{equation}

where \(\mathbf{w}_i\) is the vector of tissue volume fractions. Based on this formulation, two representative approaches have been developed. The first, partial volume pseudoinverse (PV-Pinv) \cite{deshmane2019partial}, uses a pseudoinverse projection  to estimate \(\mathbf{w}_i\). The second, PV-DM \cite{deshmane2019partial}, constructs a dictionary of signal evolutions corresponding to all possible tissue fraction combinations and applies conventional dictionary matching to recover \(\mathbf{w}_i\).

While both methods provide a mathematically grounded approach to PV estimation in MRF, they depend heavily on predefined pure-tissue relaxation parameters and manual clustering for ROI selection. These assumptions complicate robust application across large datasets containing many subjects scanned on a variety of systems. 

\section{Methodology}

As illustrated in Figure~\ref{fig:network}, DeepPV-MRF consists of two main modules: (1) a voxel-wise complex-valued multilayer perceptron (cMLP) that compresses the temporal fingerprint at each voxel into a low-dimensional latent representation, and (2) a spatial refinement module based on a U-Net architecture that predicts tissue fraction maps.
To train the network, we generated a synthetic dataset using forward Bloch simulations based on subject-derived PV maps and estimated pure-tissue relaxation parameters. The dataset incorporates radial-arm k-space undersampling and complex-valued signal formation to closely mimic real-world MRF acquisitions. We validated the proposed method using both simulated and \textit{in vivo} test data. 

The remainder of this section is organized as follows: Section~\ref{sec:network-architecture} details the network structure, Section~\ref{sec:data-generation} describes the synthetic data generation pipeline, Section~\ref{sec:train-data} describes the training data, Section~\ref{sec:test-data} outlines the synthetic testing data and \textit{in vivo} data acquisition process, Section~\ref{sec:train-net} covers the network training process, Section~\ref{sec:evaluation} presents the evaluation metrics, and Section~\ref{sec:comparison} describes the baseline comparison methods.

\subsection{Network Architecture}
\label{sec:network-architecture}
\figNetWork
The overall structure of DeepPV-MRF follows a two-stage design derived from the MRF-Mixer framework \cite{ding2025mrf, gaomrf}. The original MRF-Mixer was developed to jointly extract spatial and temporal features from aliased MRF data. In this study, we adapted and extended that architecture to explicitly target voxel-wise partial volume estimation. Figure~\ref{fig:network} provides an overview of the network, including the complex-valued temporal encoder and spatial tissue-fraction decoder.

The network consisted of two sequential modules: a voxel-wise temporal encoder and a spatial refinement module. The encoder was implemented as a complex-valued multilayer perceptron (cMLP) that independently processes each voxel’s 200-point MRF fingerprint obtained after temporal compression via singular value decomposition (SVD). It comprised four complex-valued Conv2D layers with kernel size set to \(1 \times 1\), effectively functioning as point-wise MLPs that reduce the temporal dimension to 16 latent channels. The real and imaginary parts of the resulting features were then concatenated to form a 32-channel per-voxel embedding, which was passed to the spatial decoder.

The refinement module adopted a modified U-Net to incorporate spatial context and regress tissue fraction maps. The encoder comprised four downsampling blocks with increasing channel widths \([32, 64, 128, 256]\), each consisting of a \(3 \times 3\) convolution, batch normalization, ReLU activation, and \(2 \times 2\) max pooling. The decoder was constructed symmetrically using transposed convolutions for upsampling and skip connections for spatial resolution recovery. A SoftMax activation was applied to the tissue fraction decoder output to enforce probabilistic constraints across CSF, GM, and WM fractions.

\figDataGen

For completeness, we also evaluated an auxiliary-head variant that added B0 and PD prediction branches during training. These branches were introduced as an additional supervisory signal to regularize the shared encoder, but, as shown in the ablation study, they did not improve the primary PV estimation task under the current setting. Therefore, all main simulation and \textit{in vivo} results use the no-auxiliary-head model. Overall, DeepPV-MRF enabled direct inference of PV maps from aliased MRF signals, without relying on intermediate parameter maps or dictionary matching.

\subsection{Pre-processing Synthetic Dataset}
\label{sec:data-generation}

To supervise voxel-level PV estimation, we constructed a synthetic dataset with reference PV maps and corresponding MRF signal evolutions (Figure~\ref{fig:datagen} (a)). Although specialized phantoms can be designed with known tissue mixing \cite{tang2018multicompartment}, acquiring accurate voxel-wise PV maps and pure-tissue relaxation parameters from \textit{in vivo} human subjects remains extremely challenging in practice. Therefore, we adopted a forward simulation strategy that generates synthetic MRF data from tissue compositions and relaxation values derived from real subject anatomy. This provides a practical supervisory signal for network training while maintaining precise control over the signal formation process.

Under this framework, the MRF signal \( s_i \in \mathbb{C}^t \) at voxel \( i \) was generated using the following linear mixture model:

\begin{equation}
s_i = \sum_{l=1}^{m} w_{i,l} d_l = D_{\text{sub}} \mathbf{w}_i.
\end{equation}

which is mathematically equivalent to the partial volume dictionary matching (PV-DM) formulation \cite{deshmane2019partial}. Here, \( w_{i,l} \) represents the tissue fraction of type \( l \) (CSF, GM, or WM), and \( d_l \) denotes the pure-tissue fingerprint simulated by the Bloch equation. In this study, both \( \mathbf{w}_i \) and \( D_{\text{sub}} \) were treated as known during data generation, providing precise supervision for the learning model.

We began by extracting the tissue fraction maps from high-resolution ME-MP2RAGE datasets \cite{sun2020extracting}. Specifically, FMRIB's Automated Segmentation Tool (FAST) \cite{zhang2001segmentation} was applied to subject-specific \(T_1\) maps to segment each voxel into fractional volumes of CSF (\(w_{CSF}\)), WM (\(w_{WM}\)), and GM (\(w_{GM}\)). These maps were treated as reference tissue composition maps \( \mathbf{w}_i \) at voxel \( i \) for supervision during synthetic data generation.

To estimate realistic pure-tissue T1 and T2 values, we identified high-confidence regions for each tissue type using a threshold \( p \geq 0.99 \) on the FSL FAST output:

\begin{equation}
    M_l(i) = 
    \begin{cases}
        1, & p_l(i) \geq 0.99 \\[6pt]
        0, & \text{otherwise}
    \end{cases}.
\end{equation}

These binary masks \( M_l(i) \) were then used to compute average relaxation values.

Because quantitative T2 maps were not available in the source data, approximate T2 maps were heuristically generated from the available T2* maps for MRF signal simulation. The T2* maps were first scaled using a subject-level random factor sampled between 1.2 and 1.7, followed by empirical corrections to keep the resulting values within plausible simulation ranges. These corrections included enforcing T2 $<$ T1, compressing unusually large T2 values (i.e., $> 1000~\mathrm{ms}$), applying an additional global rescaling. The resulting ranges were selected to be broadly consistent with representative 3T brain relaxation values reported in prior studies \cite{peters2007t2,wansapura1999nmr,stanisz2005t1}. This procedure was not intended as a tissue-specific physical conversion from T2* to T2, but rather as a heuristic step to introduce subject-level variability and generate approximate relaxation parameters for synthetic training-data generation.

To introduce natural variability and prevent overfitting, smooth spatial perturbations were added to \( \bar{T1}_l \) and \( \bar{T2}_l \) before Bloch simulation. These perturbations mimic intra-tissue heterogeneity and improve the realism of the synthetic signals.

Each perturbed pure-tissue relaxation pair \( (\bar{T1}_l, \bar{T2}_l) \) was passed through Bloch simulation to generate the corresponding fingerprint:

\begin{equation}
    d_l = PD \cdot \text{Bloch}(\bar{T1}_l, \bar{T2}_l, B0, FAs, TRs).
\end{equation}

where \( PD \)  is the voxel-wise proton density scaling factor at voxel \(i\), flip angles (\( FAs \)) and repetition times (\( TRs \)) are the time-varying flip angles and repetition times, and \( B0 \) is the local off-resonance frequency. The same \( B0 \) value was applied to all tissue components within each voxel as a voxel-level approximation of the dominant local field condition. This assumption was adopted to keep the multi-component signal model identifiable and computationally tractable, while acknowledging that component-specific microscopic frequency differences may exist in practice.

To emulate realistic acquisition conditions, we simulated undersampled MRF using radial k-space trajectories \cite{ding2025mrf, cloos2019rapid}, as shown in Figure~\ref{fig:datagen}(c). The image time series \( S_{\text{image}} \) was transformed into k-space using non-uniform fast Fourier transform (NUFFT):

\begin{equation}
    S_{\text{kspace\_radial}} = \text{NUFFT}(S_{\text{image}}, u_{\text{trj}}).
\end{equation}

and the aliased image series was reconstructed via:

\begin{equation}
    S_{\text{image\_aliased}} = \text{NUFFT}^{-1}(S_{\text{kspace\_radial}}, u_{\text{trj}}).
\end{equation}

The aliased MRF signal \( S_{\text{image\_aliased}} \) served as the network input.

\subsection{Synthetic Training Dataset}
\label{sec:train-data}
We applied the above procedure to ten subjects from the ME-MP2RAGE dataset \cite{sun2020extracting} for training. All imaging procedures were approved by the institutional ethics committee, and written informed consent was obtained.

To reduce memory usage and accelerate training, we applied temporal compression using SVD \cite{mcgivney2014svd}, reducing the original 1000 time points to 200 dominant components. This choice was guided by prior MRF studies showing that, for bSSFP-based dictionaries, the first 200 singular components retain nearly all dominant signal energy \cite{mcgivney2014svd}. Accordingly, we used 200 components as a robust and computationally efficient setting for evaluating the core feasibility of deep learning-based partial volume estimation from undersampled MRF signals, rather than treating SVD dimensionality itself as a primary optimization target in this study.

The SVD basis was computed once from the precomputed MRF dictionary rather than from the training or testing image data. The same fixed temporal basis was then applied to the training, synthetic testing, and \textit{in vivo} datasets. Therefore, the temporal compression step was independent of the image-level training/testing split and did not introduce information leakage from the test subjects into model training.

The final training dataset contained 1200 paired samples of size \(192 \times 192\), each consisting of compressed aliased MRF signals as input and the corresponding CSF, WM, and GM tissue fraction maps as labels. To enhance spatial diversity and improve generalization, we applied random cropping during training, as illustrated in Figure~\ref{fig:datagen}(d). Specifically, each original sample was randomly cropped five times, resulting in a total of 30,000 training samples of size \(64 \times 64 \times 200\).

\subsection{Testing Dataset}
\label{sec:test-data}

To evaluate model performance under controlled conditions, we first constructed a synthetic test set using three subjects from the ME-MP2RAGE dataset \cite{sun2020extracting}, which were not used during training. For each subject, we selected slices primarily located in the central brain region to ensure anatomical diversity. A total of 120 synthetic test samples were generated, each of size \(192 \times 192\). These samples were used exclusively during inference to provide quantitative comparisons of reconstruction accuracy across different models.

In addition to the synthetic test data, \textit{in vivo} datasets were acquired from three healthy volunteers (sub 01, sub 02, sub 03). The sub 01 dataset was acquired on a 3~T Siemens Prisma scanner, whereas the sub 02 and sub 03 datasets were acquired after the scanner system was upgraded to a 3~T Siemens MAGNETOM Cima.X Fit platform. Each volunteer was scanned using single-shot, three-shot, and six-shot MRF acquisitions, resulting in nine \textit{in vivo} datasets in total. Imaging was performed across 32 slices using a 1000-point IR-bSSFP sequence with varying radial undersampling strategies, an in-plane resolution of 1~mm, and a slice thickness of 3~mm. Written informed consent was obtained from all volunteers prior to participation. The trained models were directly applied to these datasets without retraining or fine-tuning, and the results were used to evaluate cross-shot consistency and multi-volunteer \textit{in vivo} robustness across the expanded scanner/protocol setting.

\subsection{Network Training}
\label{sec:train-net}
The DeepPV-MRF network was trained using the Adam optimizer on an NVIDIA GeForce RTX 5090 GPU for 100 epochs. All model parameters were initialized from a normal distribution with a mean of \(0\) and a standard deviation of \(0.01\), and the full training process required approximately one hour under the current implementation settings. After training, computational profiling of the final model showed that it contained 7.14 million trainable parameters and required approximately \(201.81\) GFLOPs for one \(192 \times 192\) slice. On the same GPU, the inference time was \(3.06\) ms per slice, with peak GPU memory usage of \(0.243\) GB allocated and \(0.303\) GB reserved.

The network was trained by minimizing a tissue fraction estimation loss. This loss contains two terms: an \(L_1\) loss for voxel-wise tissue fraction accuracy and a spatial gradient consistency loss for preserving tissue boundaries:

\begin{equation}
\mathcal{L}_{\text{PV}} = \mathcal{L}_{\text{L1}}(\hat{W}, W) + \lambda_{\text{spatial}} \cdot \mathcal{L}_{\text{spatial}}(\hat{W}, W),
\end{equation}

where \(\hat{W}\) and \(W\) denote the predicted and reference tissue fraction maps, respectively.

The \(L_1\) loss promotes voxel-wise accuracy and is defined as:
\begin{equation}
\mathcal{L}_{\text{L1}}(\hat{W}, W) = \frac{1}{N} \sum_{i=1}^N \left| \hat{W}_i - W_i \right|.
\end{equation}

To enforce spatial smoothness, we used a first-order gradient consistency loss:
\begin{align}
X_{x, i,j} &= \left| \nabla_x \hat{W}_{i,j} - \nabla_x W_{i,j} \right|, \\
X_{y, i,j} &= \left| \nabla_y \hat{W}_{i,j} - \nabla_y W_{i,j} \right|, \\
\mathcal{L}_{\text{spatial}}(\hat{W}, W) &= \frac{1}{N} \sum_{i,j} \left( X_{x, i,j} + X_{y, i,j} \right),
\end{align}
where \(\nabla_x\) and \(\nabla_y\) denote finite difference approximations of spatial gradients in the horizontal and vertical directions, respectively.

The hyperparameter $\lambda_{\text{spatial}}$ controls the trade-off between zeroth-order voxel-wise fidelity and first-order spatial gradient consistency. In the main experiments, we used $\lambda_{\text{spatial}} = 0.05$ as a default setting because it provided stable training and satisfactory reconstruction quality during model development. As shown later in the ablation study, performance remained relatively stable across a range of values.

For the auxiliary-head ablation only, the objective was extended as:
\begin{equation}
\mathcal{L}_{\text{Total}}^{\text{aux}} = \mathcal{L}_{\text{PV}} + \mathcal{L}_{B0} + \mathcal{L}_{PD},
\end{equation}
where \(\mathcal{L}_{B0} = \text{MSE}(\hat{B0}, B0)\) and \(\mathcal{L}_{PD} = \text{MSE}(\hat{PD}, PD)\) represent voxel-wise mean squared errors between predicted and reference B0 and PD maps, respectively.

\subsection{Evaluation Metrics}
\label{sec:evaluation}
To evaluate estimation performance, we employed voxel-wise metrics to reflect numerical accuracy. These quantitative voxel-wise metrics were computed for the synthetic test data, where reference tissue fraction maps were available. For the \textit{in vivo} data, no voxel-wise reference PV maps were available; therefore, metrics requiring reference tissue fractions were not computed in that setting. Instead, an expert manually delineated the brain contour in ImageJ \cite{schneider2012nih} to generate a brain mask, which was used to restrict the analysis and visualization to the brain region.

\subsubsection{Voxel-wise Accuracy Metrics}
We used the following voxel-wise metrics to quantify the accuracy of the estimated tissue fraction maps on the synthetic test set: root mean square error (RMSE), tissue-specific mean absolute error (MAE), peak signal-to-noise ratio (PSNR), structural similarity index (SSIM), and weighted Pearson correlation coefficient (\(r_{\text{weighted}}\)). These metrics collectively assess both numerical fidelity (RMSE, MAE, PSNR) and structural similarity (SSIM). Because these quantities require voxel-wise reference tissue fractions, they were not computed for the \textit{in vivo} data.

For a given tissue \(t \in \{\mathrm{CSF}, \mathrm{WM}, \mathrm{GM}\}\), let \(w_t(i)\) denote the reference fraction of tissue \(t\) at voxel \(i\), and let \(y_{\mathrm{pred},t,i}\) and \(y_{\mathrm{true},t,i}\) denote the predicted and reference values, respectively. The tissue-specific partial-volume mask was defined as
\begin{equation}
    \Omega_{PV,t} = \{\, i \mid 0 < w_t(i) < 1,\; i \in \Omega_{\mathrm{brain}} \,\},
\end{equation}
where \(\Omega_{\mathrm{brain}}\) denotes the brain mask, and \(N_{PV,t}\) is the number of voxels in \(\Omega_{PV,t}\).

\begin{itemize}
    \item \textbf{Tissue-specific Mean Absolute Error (MAE)}: The tissue-specific MAE was evaluated on the corresponding partial-volume region to assess accuracy in transition areas:
    \begin{equation}
        \mathrm{MAE}_{t} = \frac{1}{N_{PV,t}} \sum_{i \in \Omega_{PV,t}} \left| y_{\mathrm{pred},t,i} - y_{\mathrm{true},t,i} \right|.
    \end{equation}

    In addition to the PV-restricted tissue-specific MAE, we also report whole-brain metrics computed over all voxels within the brain mask to facilitate comparison with related work in Supplementary Table~2.

    \item \textbf{Weighted Pearson Correlation (\(r_{\text{weighted}}\))}: We computed a tissue-specific weighted Pearson correlation on \(\Omega_{PV,t}\), with weights defined by
    \begin{equation}
        \omega_{t,i} = w_t(i)\bigl(1 - w_t(i)\bigr).
    \end{equation}
    The weighted correlation was then computed as
    \begin{align}
        \tilde{y}_{t,i} &= y_{\mathrm{true},t,i} - \bar{y}_{\mathrm{true},t}, \\
        \hat{y}_{t,i} &= y_{\mathrm{pred},t,i} - \bar{y}_{\mathrm{pred},t}, \\
        r_{\mathrm{weighted},t} &=
        \frac{\sum_{i \in \Omega_{PV,t}} \omega_{t,i}\,\tilde{y}_{t,i}\hat{y}_{t,i}}
        {\sqrt{\sum_{i \in \Omega_{PV,t}} \omega_{t,i}\tilde{y}_{t,i}^2}
        \sqrt{\sum_{i \in \Omega_{PV,t}} \omega_{t,i}\hat{y}_{t,i}^2}}.
    \end{align}
\end{itemize}

In addition to quantitative evaluations, we generated voxel-wise error maps for qualitative analysis, enabling visual inspection of regional estimation errors and anatomical plausibility.

\subsection{Comparison with Other Methods}
\label{sec:comparison}

To comprehensively assess the performance of the proposed method, we conducted comparative analyses against several established approaches, including PV-Pinv \cite{deshmane2019partial}, and PV-DM \cite{deshmane2019partial}. Because these competing methods operate on different signal representations and require different prior information, we explicitly distinguish below between matched and unmatched comparison conditions.

In the simulation experiments, DeepPV-MRF, PV-Pinv, and PV-DM were evaluated using the same synthetic MRF inputs generated from the same underlying tissue composition and relaxation parameter framework. For PV-Pinv and PV-DM, the required dictionaries \( D_{sub} \) and \( D_{PV} \) were constructed from the same reference tissue parameters used in the simulation setup. Therefore, the simulation comparison among these MRF-based methods was performed under closely matched input conditions and parameter assumptions.

For the \textit{in vivo} experiment, some differences in preconditions were unavoidable because the methods operate in different input domains. Following the original PV-DM study \cite{deshmane2019partial}, we first generated T1 and T2 maps through dictionary matching and then applied k-means clustering with \( k = 11 \) to select representative T1 and T2 values for constructing \( D_{sub} \) and \( D_{PV} \) for PV-Pinv and PV-DM. The value \(k=11\) was adopted from the original PV-DM implementation rather than tuned in this study. The resulting cluster centers were used as representative T1/T2 parameter candidates spanning the observed tissue distribution, not as eleven final tissue classes. Candidate centers corresponding to CSF, GM, and WM were then used to construct the three-component PV dictionary for tissue fraction estimation. In contrast, DeepPV-MRF directly used aliased MRF time-series signals as input. 

Accordingly, the comparison with PV-Pinv and PV-DM can be regarded as an MRF-based methodological comparison under closely aligned simulation settings. We have made these distinctions explicit to clarify both the fairness and the limitations of the comparative evaluation.

\figSimulation
\section{Results}

\subsection{Simulation Results}

\tableAllShots

\figScatterPlot
Figure~\ref{fig:simulation} shows a qualitative comparison of partial volume estimation results from PV-DM, PV-Pinv, and DeepPV-MRF across single-shot, three-shot, and six-shot acquisitions. For each method, CSF, WM, and GM tissue fraction maps are shown along with their corresponding difference maps against the reference labels. For all shot settings, DeepPV-MRF consistently showed the best image contrasts compared with the reference labels, especially around tissue boundaries and regions with partial volume effects. The improvement is most noticeable in the single-shot case, where PV-DM and PV-Pinv showed clear structural errors, as was confirmed in Table~\ref{tab:pv_comparison_combined} which summarized the quantitative comparisons of different methods. Particularly, DeepPV-MRF outperformed both PV-DM and PV-Pinv on all metrics. For example, under the single-shot setting, DeepPV-MRF achieved lower MAE values ($41.76 \times 10^{-3}$ for CSF, $74.39 \times 10^{-3}$ for WM, and $75.69 \times 10^{-3}$ for GM) and higher SSIM scores (0.960, 0.916, and 0.922 respectively). Figure~\ref{fig:scatterplot} shows scatter plots of predicted PV values of DeepPV-MRF against the reference labels. As the number of shots increases from 1 to 6, the data points become more concentrated around the diagonal line, indicating better agreement with the reference labels. Correlation values and trend line slopes also increased accordingly. For the 6-shot case, DeepPV-MRF achieved MAE values of $22.93 \times 10^{-3}$ (CSF), $43.16 \times 10^{-3}$ (WM), and $43.96 \times 10^{-3}$ (GM), with PSNR values reaching 35.12~dB (CSF), 29.35~dB (WM), and 28.22~dB (GM). These results confirm that DeepPV-MRF provides more accurate and stable partial volume estimation under different acquisition settings, especially in low-shot scenarios.

To further assess statistical robustness, paired non-parametric statistical tests were performed across the synthetic test slices. The Holm-adjusted Friedman tests and post-hoc paired Wilcoxon signed-rank tests are reported in Supplementary Table~1. These results support the quantitative trends in Table~\ref{tab:pv_comparison_combined}, showing consistent improvements of DeepPV-MRF over PV-DM and PV-Pinv across shot settings.

\figInvivoOne

\subsection{\textit{in vivo} results}

Figure~\ref{fig:invivo1} shows the \textit{in vivo} tissue fraction maps for PV-DM, PV-Pinv, and DeepPV-MRF under single-shot, three-shot, and six-shot acquisitions from sub 01. CSF, WM, and GM maps are displayed in separate columns. Qualitatively, PV-DM produced relatively stable CSF maps across shot settings, but the WM and GM maps exhibited visible noise and aliasing, particularly in the single-shot and three-shot acquisitions. PV-Pinv generated smoother CSF maps, yet the WM and GM maps appeared spatially blurred and showed reduced structural detail across all acquisition settings. In contrast, DeepPV-MRF produced tissue fraction maps with more consistent intensity distributions and clearer anatomical structures for CSF, WM, and GM under all three shot settings.

\figInvivoNew

Figure~\ref{fig:invivonew} shows the corresponding DeepPV-MRF results from the two additional volunteers, sub 02 and sub 03, to further evaluate the model performance under \textit{in vivo} acquisition conditions. As shown in Figure~\ref{fig:invivonew}, the proposed method produced visually consistent CSF, WM, and GM fraction maps across both subjects and different acquisition settings, including single-shot, three-shot, and six-shot scans. The main anatomical structures, including the ventricles, cortical GM, and deep WM regions, were well preserved in all three tissue fraction maps. Compared with the single-shot results, the three-shot and six-shot acquisitions showed improved spatial smoothness and clearer tissue boundaries, while maintaining a similar overall tissue distribution pattern.

To complement this qualitative assessment, Table~\ref{tab:invivo_cross_shot_consistency} summarizes the cross-shot consistency of DeepPV-MRF across the three \textit{in vivo} volunteers by using each subject's six-shot prediction as an internal high-shot reference. Across all nine subject-tissue pairs, the three-shot predictions showed lower MAE than the corresponding single-shot predictions, with an average MAE reduction of 15.3\%, supporting improved agreement with the six-shot reference as the acquisition increased from single-shot to three-shot. These results suggest that DeepPV-MRF can provide stable tissue fraction estimation across different subjects, anatomical instances, and the scanner/protocol settings included in this \textit{in vivo} evaluation.

\tableInvivoCrossShotConsistency

\figInvivoSNR

To further assess robustness under degraded acquisition conditions, we performed an additional \textit{in vivo} noise-sensitivity analysis for the 1-shot setting across the three volunteers by perturbing the input MRF signals with complex Gaussian noise over a range of SNR levels. As shown in Figure~\ref{fig:invivosnr}, the MAE between noisy-input predictions and the corresponding clean-input predictions decreased monotonically as SNR increased for CSF, WM, and GM in all three volunteers, indicating that the proposed framework remained stable under progressively noisier inputs.

\tableAblation

\subsection{Ablation Study}
To quantify the contribution of the main architectural and loss design choices, we conducted ablation experiments on the one-shot external evaluation set, including removal of the spatial refinement module, addition of auxiliary B0/PD heads to the final model, removal of the spatial gradient loss, variation of the latent dimension, and a sweep of the spatial regularization weight $\lambda_{\text{spatial}}$. The results are summarized in Table~\ref{tab:ablation}.

Among the core ablations, removing the spatial refinement module led to the largest performance drop, confirming that spatial context is essential for accurate PV estimation. The temporal encoder alone increased the macro tissue-wise MAE from 63.95 to 89.54 and reduced average PSNR/SSIM from 26.48/0.933 to 23.95/0.890, indicating that the main performance gain arises from the joint temporal-spatial design rather than from voxel-wise temporal encoding alone.

The final full model does not include auxiliary B0/PD heads. Adding these heads slightly degraded PV estimation performance in the current setting, increasing the macro tissue-wise MAE from 63.95 to 66.55 and reducing average PSNR/SSIM from 26.48/0.933 to 26.14/0.922. This suggests that, although these auxiliary branches were originally introduced to regularize the shared encoder, their supervision was not optimally aligned with the primary PV estimation objective and may have introduced mild optimization competition\cite{ruder2017overview}. Removing the spatial gradient loss also caused a modest degradation relative to the full model, increasing the macro tissue-wise MAE to 67.23. In the latent-dimension sweep, latent dimensions of 8 and 16 produced comparable performance, with the latent dimension of 8 yielding the lowest macro tissue-wise MAE on this one-shot external set, whereas larger latent dimensions of 32, 64, and 128 degraded the WM and GM metrics. For the spatial regularization sweep, performance remained relatively stable across a range of nonzero $\lambda_{\text{spatial}}$ values, and the default setting of $\lambda_{\text{spatial}} = 0.05$ provided the best macro-averaged performance among the tested nonzero weights.

\section{Discussion}

This study presents DeepPV-MRF, a deep learning framework for direct partial volume estimation from undersampled MRF signals. Unlike conventional methods such as PV-DM, PV-Pinv, and FSL FAST, the proposed approach avoids strong reliance on hand-crafted priors and multi-step intermediate processing. By directly mapping aliased MRF time-series signals to tissue fraction maps, DeepPV-MRF simplifies the workflow and reduces the risk of error propagation from intermediate reconstructions or model assumptions.

The performance gains of DeepPV-MRF appear to arise from two main factors. First, the simulation-based training pipeline combines subject-derived tissue property maps and volumetric fraction maps with realistic radial undersampling, allowing the network to learn tissue-specific signal behavior under practical aliasing conditions. Second, the network architecture combines voxel-wise temporal encoding with spatial refinement, which is particularly important in regions with complex tissue mixing. The ablation study supports this interpretation: removing the spatial refinement module caused the largest performance drop, whereas adding auxiliary B0/PD heads and removing the spatial gradient loss had smaller effects on the final PV estimation accuracy.

The simulation results demonstrated clear advantages of DeepPV-MRF over PV-DM and PV-Pinv across all shot settings, with particularly strong performance under the more challenging single-shot condition. The \textit{in vivo} results further supported these findings, as DeepPV-MRF produced tissue fraction maps with smoother transitions and better anatomical continuity, especially in GM and WM regions.

The multi-volunteer \textit{in vivo} cross-shot consistency analysis provided additional support for this interpretation. Because voxel-wise \textit{in vivo} tissue-fraction ground truth was unavailable, the six-shot prediction was used only as an internal high-shot reference rather than as an accuracy standard. Across all nine subject-tissue pairs, the three-shot estimates showed lower MAE relative to the six-shot reference than the single-shot estimates, with an average MAE reduction of 15.3\%. This trend indicates that increasing the number of shots improved agreement with the high-shot reference across subjects and tissues, while preserving the visually observed anatomical structures in the estimated fraction maps.

\figMixer

The comparison between composite multi-component T1/T2 maps and single-component MRF-Mixer estimates further highlights the value of explicit partial volume modeling. As shown in Figure~\ref{fig:testmixer}, the largest discrepancies were observed near tissue interfaces, where partial volume effects are expected to be strongest. This suggests that mixed-tissue MRF voxels are not always well represented by single-component relaxation estimates, and that explicit PV estimation may support future frameworks for joint estimation of tissue fractions and intrinsic tissue parameters.

From a practical perspective, the inference benchmark reported in the Network Training section indicates that DeepPV-MRF can generate tissue-fraction maps with low per-slice latency and modest GPU memory usage. In addition, the three-volunteer \textit{in vivo} noise perturbation analysis showed monotonically decreasing prediction deviation as SNR increased, providing preliminary evidence of robustness to moderate noise variation.

Several limitations should be acknowledged. Although the \textit{in vivo} evaluation included three healthy volunteers acquired from two different scanner platforms (one from 3T Prisma and two from 3T Cima.x Fit), the validation does not yet establish full generalizability across sites, vendors, acquisition protocols, or pathological populations. In addition, the supervisory tissue fraction maps were derived from FSL FAST on high-resolution anatomical images, so they should be regarded as surrogate references rather than absolute ground truth, and some FAST-related bias may therefore be inherited by the model. The synthetic signal model also assumes linear combinations of pure-tissue fingerprints \cite{deshmane2019partial}, which may not fully capture effects such as magnetization transfer, exchange, or other multi-compartment nonlinearities in real tissues. Future work should therefore include broader validation across scanners, protocols, B1 conditions, motion levels, and pathological tissues, together with more independent reference standards such as phantoms or expert-curated annotations.

\section{Conclusion}

In summary, we proposed a deep learning-based method for partial volume estimation that directly maps MRF signals to tissue fraction maps in an end-to-end approach. By integrating subject-specific simulation data with radial arm undersampling and employing a spatial-temporal cascaded network architecture, the method achieves improved accuracy and robustness compared to traditional approaches such as PV-DM and PV-Pinv. These advantages simplify the processing pipeline and improve its suitability for practical applications. Future work will focus on validating the method across more diverse anatomical regions and clinical conditions, exploring self-supervised or unsupervised training strategies, and extending the approach to additional acquisition sequences.

\section*{References}
\bibliography{reference}      

\bibliographystyle{medphy}    

\clearpage
\section*{Supplementary Material}
\setcounter{table}{0}
\renewcommand{\tablename}{Supplementary Table}
\renewcommand{\theHtable}{supp.\arabic{table}}
\begingroup
\renewenvironment{table*}[1][]{\begin{table}[H]}{\end{table}}
\begin{table*}[!t]
\centering
\caption{
Paired statistical analysis using the same simulation slices as in main manuscript Table~1.
For each shot setting, tissue, and metric, we report the Holm-adjusted Friedman $p$-value across the three methods and the median paired difference ($\Delta$) of DeepPV-MRF relative to each baseline.
For MAE, negative values indicate lower error; for PSNR, SSIM, and $r_{\mathrm{weighted}}$, positive values indicate improved performance.
Across all post-hoc paired Wilcoxon signed-rank tests, the Holm-adjusted $p$-value was $6.83\times10^{-20}$.
}
\label{tab:supp_significance}
\footnotesize
\resizebox{\textwidth}{!}{%
\begin{tabular}{lllccc}
\toprule
Shot setting & Tissue & Metric & Friedman $p_{\mathrm{Holm}}$ & $\Delta$ vs PV-DM & $\Delta$ vs PV-Pinv \\
\midrule
\multirow{12}{*}{Single-shot}
& \multirow{4}{*}{CSF} & MAE ($\times 10^3$) & $2.78\times10^{-52}$ & $-0.094$ & $-0.153$ \\
& & PSNR & $2.78\times10^{-52}$ & $11.434$ & $16.549$ \\
& & SSIM & $1.37\times10^{-52}$ & $0.332$ & $0.554$ \\
& & $r_{\mathrm{weighted}}$ & $1.37\times10^{-52}$ & $0.259$ & $0.477$ \\
\cmidrule(lr){2-6}
& \multirow{4}{*}{WM} & MAE ($\times 10^3$) & $6.01\times10^{-51}$ & $-0.165$ & $-0.209$ \\
& & PSNR & $1.93\times10^{-48}$ & $12.611$ & $13.350$ \\
& & SSIM & $1.37\times10^{-52}$ & $0.379$ & $0.467$ \\
& & $r_{\mathrm{weighted}}$ & $3.85\times10^{-48}$ & $0.424$ & $0.509$ \\
\cmidrule(lr){2-6}
& \multirow{4}{*}{GM} & MAE ($\times 10^3$) & $2.91\times10^{-42}$ & $-0.233$ & $-0.181$ \\
& & PSNR & $4.03\times10^{-42}$ & $12.749$ & $12.262$ \\
& & SSIM & $4.33\times10^{-44}$ & $0.414$ & $0.430$ \\
& & $r_{\mathrm{weighted}}$ & $4.03\times10^{-42}$ & $0.582$ & $0.528$ \\
\midrule
\multirow{12}{*}{Three-shot}
& \multirow{4}{*}{CSF} & MAE ($\times 10^3$) & $1.43\times10^{-51}$ & $-0.090$ & $-0.150$ \\
& & PSNR & $6.01\times10^{-52}$ & $13.099$ & $18.162$ \\
& & SSIM & $1.37\times10^{-52}$ & $0.302$ & $0.509$ \\
& & $r_{\mathrm{weighted}}$ & $1.43\times10^{-51}$ & $0.218$ & $0.402$ \\
\cmidrule(lr){2-6}
& \multirow{4}{*}{WM} & MAE ($\times 10^3$) & $2.78\times10^{-52}$ & $-0.157$ & $-0.231$ \\
& & PSNR & $2.75\times10^{-51}$ & $14.640$ & $16.480$ \\
& & SSIM & $1.37\times10^{-52}$ & $0.363$ & $0.486$ \\
& & $r_{\mathrm{weighted}}$ & $2.78\times10^{-52}$ & $0.391$ & $0.544$ \\
\cmidrule(lr){2-6}
& \multirow{4}{*}{GM} & MAE ($\times 10^3$) & $2.62\times10^{-42}$ & $-0.221$ & $-0.235$ \\
& & PSNR & $3.65\times10^{-43}$ & $14.765$ & $15.502$ \\
& & SSIM & $1.37\times10^{-52}$ & $0.371$ & $0.514$ \\
& & $r_{\mathrm{weighted}}$ & $1.92\times10^{-49}$ & $0.557$ & $0.691$ \\
\midrule
\multirow{12}{*}{Six-shot}
& \multirow{4}{*}{CSF} & MAE ($\times 10^3$) & $1.43\times10^{-51}$ & $-0.082$ & $-0.149$ \\
& & PSNR & $1.43\times10^{-51}$ & $13.155$ & $18.550$ \\
& & SSIM & $1.37\times10^{-52}$ & $0.282$ & $0.508$ \\
& & $r_{\mathrm{weighted}}$ & $1.43\times10^{-51}$ & $0.197$ & $0.389$ \\
\cmidrule(lr){2-6}
& \multirow{4}{*}{WM} & MAE ($\times 10^3$) & $6.01\times10^{-52}$ & $-0.152$ & $-0.238$ \\
& & PSNR & $2.75\times10^{-51}$ & $15.773$ & $17.881$ \\
& & SSIM & $1.37\times10^{-52}$ & $0.347$ & $0.479$ \\
& & $r_{\mathrm{weighted}}$ & $2.78\times10^{-52}$ & $0.363$ & $0.545$ \\
\cmidrule(lr){2-6}
& \multirow{4}{*}{GM} & MAE ($\times 10^3$) & $2.69\times10^{-43}$ & $-0.206$ & $-0.241$ \\
& & PSNR & $1.13\times10^{-43}$ & $15.351$ & $16.838$ \\
& & SSIM & $1.37\times10^{-52}$ & $0.357$ & $0.524$ \\
& & $r_{\mathrm{weighted}}$ & $4.10\times10^{-49}$ & $0.536$ & $0.701$ \\
\bottomrule
\end{tabular}%
}
\end{table*}

\begin{table*}[!t]
\centering
\caption{Whole-brain error metrics for the DeepPV-MRF model under different shot settings. Metrics were computed over all voxels within the brain mask across the same 123 external synthetic test slices used for the main simulation evaluation. Values are reported as mean $\pm$ standard deviation.}
\label{tab:supp_wholebrain_metrics}
\footnotesize
\resizebox{0.80\textwidth}{!}{%
\begin{tabular}{llcc}
\toprule
Shot setting & Tissue & MAE ($\times10^3$) ($\downarrow$) & RMSE ($\times10^3$) ($\downarrow$) \\
\midrule
\multirow{3}{*}{Single-shot} & CSF & 19.25 $\pm$ 3.62 & 43.35 $\pm$ 7.50 \\
& WM  & 38.98 $\pm$ 7.53 & 83.00 $\pm$ 13.41 \\
& GM  & 55.34 $\pm$ 10.12 & 95.24 $\pm$ 15.34 \\
\midrule
\multirow{3}{*}{Three-shot} & CSF & 12.56 $\pm$ 3.26 & 28.90 $\pm$ 7.57 \\
& WM  & 29.23 $\pm$ 5.24 & 59.49 $\pm$ 9.15 \\
& GM  & 39.88 $\pm$ 8.06 & 69.26 $\pm$ 12.51 \\
\midrule
\multirow{3}{*}{Six-shot} & CSF & 11.53 $\pm$ 2.59 & 26.66 $\pm$ 6.30 \\
& WM  & 24.49 $\pm$ 4.47 & 49.19 $\pm$ 8.25 \\
& GM  & 33.71 $\pm$ 6.11 & 57.57 $\pm$ 9.55 \\
\bottomrule
\end{tabular}%
}
\end{table*}

\endgroup
\end{document}